\documentclass[preprint,authoryear,12pt]{article}
\usepackage{amssymb}
\usepackage{graphicx}
\usepackage{booktabs}
\usepackage{multirow}
\usepackage{siunitx}
\usepackage{cite}
\usepackage{lineno}

\begin{document}

\begin{center}
\LARGE{\textbf{Analysis of time-profiles with in-beam PET monitoring in charged particle therapy}}
\end{center}
\vspace{0.2cm}

\normalsize{\noindent A.C. Kraan$^{a,*}$, S. Muraro$^a$, G. Battistoni$^b$, N. Belcari$^{a,c}$, M.G. Bisogni$^{a,c}$, N. Camarlinghi$^{a,c}$, A. Del Guerra$^{a,c}$, A. Ferrari$^d$, R. Kopec$^e$, D. Krzempek$^e$, M. Morrocchi$^{a,c}$, P. Olko$^e$, P. Sala$^b$, K. Skowro\`nska$^e$, G. Sportelli$^{a,c}$, A. Topi$^{a,f}$, V. Rosso$^{a,c}$}

\vspace{0.6cm}

\footnotesize{
a  Istituto Nazionale di Fisica Nucleare, Sezione di Pisa, Italy;  

b  Istituto Nazionale di Fisica Nucleare, Sezione di Milano, Italy;      

c  Department of Physics, University of Pisa, Italy;      

d  CERN, Geneva, Switzerland ;

e  Institute of Nuclear Physics Polish Academy of Sciences, Krakow, Poland   

f  Department of Physical Sciences, Earth and Environment, University of Siena, Italy.

\vspace{0.2cm}
* Corresponding author, email: aafke.kraan@pi.infn.it
}
\vspace{0.4cm}
\begin{abstract}
\textbf{Background:} Treatment verification with PET imaging in charged particle therapy is conventionally done by comparing measurements of spatial distributions with Monte Carlo (MC) predictions. However, decay curves can provide additional independent information about the treatment and the irradiated tissue. Most studies performed so far focus on long time intervals. Here we investigate the reliability of MC predictions of space and time (decay rate) profiles shortly after irradiation, and we show how the decay rates can give an indication about the elements of which the phantom is made up.

\textbf{Methods and Materials:} Various phantoms were irradiated in clinical and near-clinical conditions at the Cyclotron Centre of the Bronowice proton therapy centre. PET data were acquired with a planar 16x16 cm$^2$ PET system. MC simulations of particle interactions and photon propagation in the phantoms were performed using the FLUKA code. The analysis included a comparison between experimental data and MC simulations of space and time profiles, as well as a fitting procedure to obtain the various isotope contributions in the phantoms.

\textbf{Results and conclusions:} There was a good agreement between data and MC predictions in 1-dimensional space and decay rate distributions. The fractions of $^{11}$C, $^{15}$O and $^{10}$C that were obtained by fitting the decay rates with multiple simple exponentials generally agreed well with the MC expectations. We found a small excess of $^{10}$C in data compared to what was predicted in MC, which was clear especially in the PE phantom.

\end{abstract}

\newpage
\normalsize
\section{Introduction}
In charged particle therapy cancerous tissue is irradiated with charged particles. The quality of charged particle therapy treatments depends on the ability to predict and achieve a given particle range in the patient. In-beam PET (Fig. 1) is a non-invasive method that can be used to estimate the particle range during or shortly after patient irradiation (``in-beam''). PET monitoring is based on the detection of $\beta^+$ emitters (predominantly $^{11}$C and $^{15}$O) produced in the patient as a result of nuclear interactions of charged hadrons with tissue~\cite{enghardt2004b,iseki2004}. Reviews about PET as monitoring tool in charged paricle therapy can be found for instance in~\cite{zhu2013, nishio2010,knopf,parodi,kraan}. Treatment verification is commonly done by comparing  measured and pre-calculated MC distributions in space. However, the decay curve of the activated material can provide additional information about the treatment and about the patient, because its shape depends on the decaying isotopes and thus on the irradiated tissue~\cite{grogg,cho2013,akagi2013, dendooven2015, cambria, matsu2016,buitenhuis2017, brombal2017}. Mapping of $^{15}$O can for instance be highly useful to investigate biological washout models and perfusion in patients~\cite{grogg}. Moreover, from the relative radioisotope fractions it is possible to calculate the elemental composition of the irradiated tissue, which is useful to detect changes of oxygenation in tumors and radiation induced necrosis~\cite{cho2013}. Finally, time profiles are relevant to validate the low-energy interaction nuclear physics methods in MC codes. 

\begin{figure}[b!]
\includegraphics[width=0.8\textwidth]{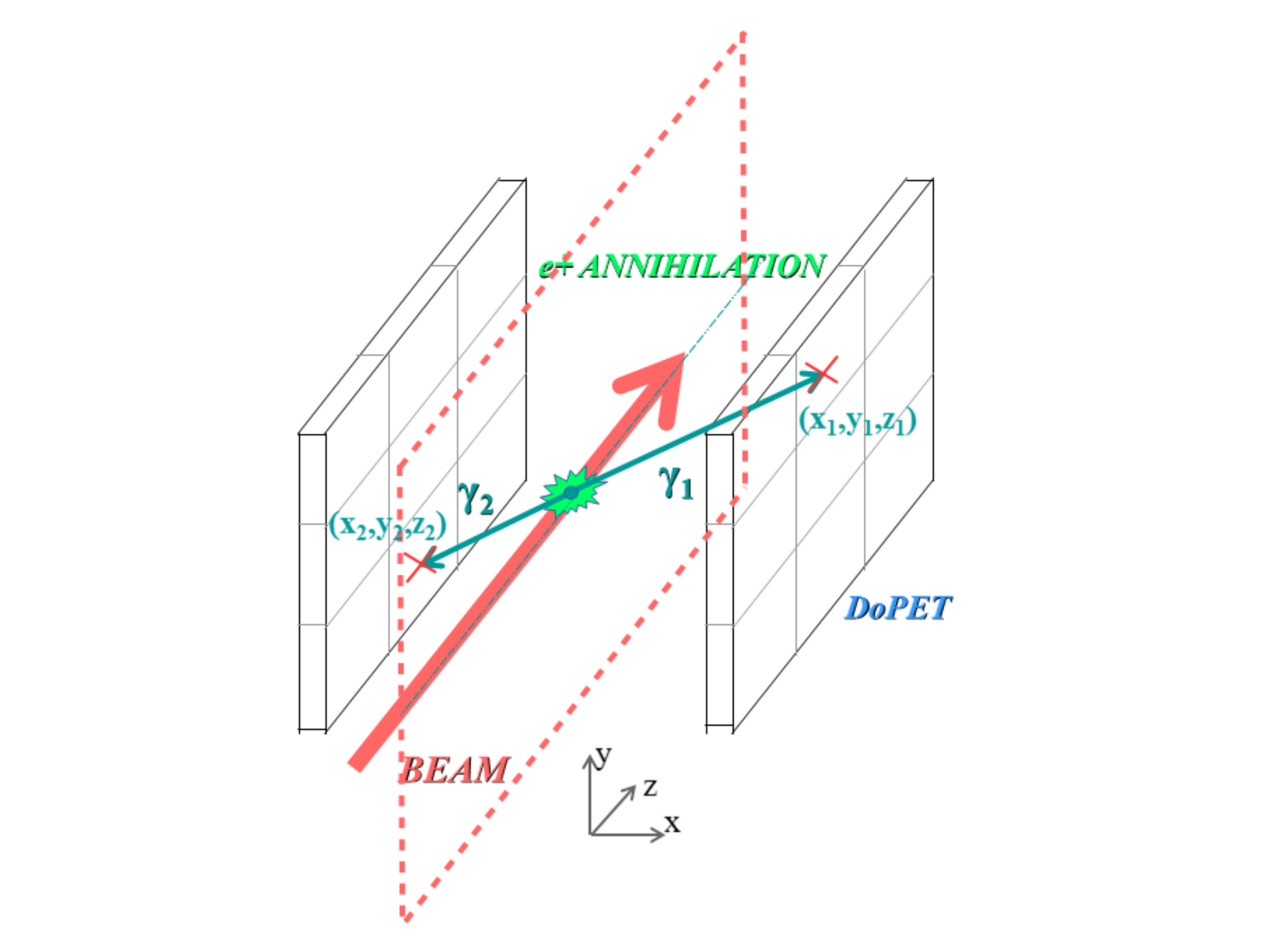}
\caption{\footnotesize{PET imaging in charged particle therapy. }}
\label{fig1}
\end{figure}
Information about the decaying isotopes is particularly useful if it can be obtained shortly after irradiation. Still, much literature focuses on long-time intervals~\cite{grogg,cho2013, akagi2013,matsu2016}, loosing the valuable data that can be acquired just after end of irradiation. Concerning short time intervals, Buitenhuis et. al.~\cite{buitenhuis2017} focus on PET imaging during irradiation, acquiring the $^{12}$N signal, but not in the context of tissue composition. In the current study we focus on short time intervals dominated by $^{11}$C, $^{15}$O, and $^{10}$C, similar to works by Cambraia Lopes et. al.~\cite{cambria}, Matsushita et. al~\cite{matsu2016}, and previous research in our group~\cite{brombal2017}. However, the goal of the current work is to extract the relative fractions of $^{15}$O, $^{11}$C and $^{10}$C decaying in the phantoms using even shorter time intervals than those previously reported, and under different irradiation conditions. Moreover we include also an inhomogeneous phantom and show an example of how the different isotopes can be mapped. Finally, we use a different MC release. Thus, this work complements the available literature on the subject. 


\section{Methods and Materials:}
We used three homogeneous phantoms (PMMA, high density PE and Water) and one inhomogeneous phantom (Zebra: PMMA and high density PE) of 5 x 5 x 15 cm$^3$. 
The Zebra phantom consisted of consecutive layers of 2 cm PMMA, 2 cm high-density PE, 2 cm PMMA, 2 cm high-density PE and 7 cm PMMA. In the following we denote high density PE simply as PE. These phantoms were irradiated during 5 s with single pencil beams ($10^{10}$ protons, FWHM=10.7 mm) at the Cyclotron Centre of the Bronowice proton therapy centre in Krakow, Poland. The beam energy was 130 MeV.

The PET system used for data acquisition was a planar 16x16 cm$^2$ PET system (DoPET~\cite{rosso}) based on LYSO crystals and Hamamatsu H8500 position sensitive photo-multipliers (Fig. 2).
\begin{figure}[t!]
\includegraphics[width=0.65\textwidth]{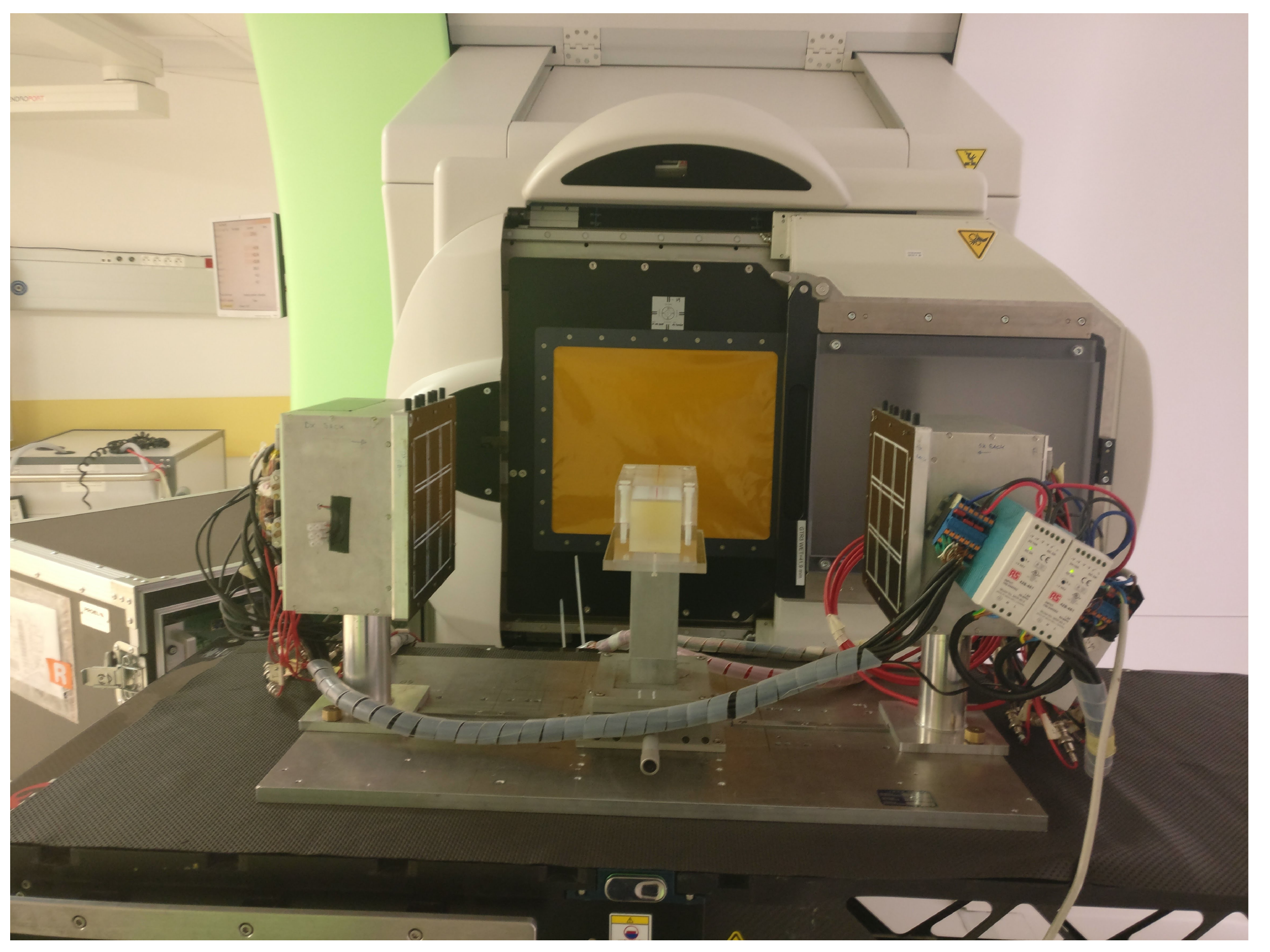}
\caption{\footnotesize{The PET system acquiring data at the Bronowice proton therapy centre.}}
\label{fig2}
\end{figure}

As image reconstruction method we used a non-conventional method (``Straightforward Reconstruction Approach'', SRA~\cite{muraro}). The annihilation position and time of each event was calculated as follows. The $x$-coordinate was given by the known x-position of the beam, which was here the center of the two detector planes. The $y$ and $z$-coordinates of each annihilation were be obtained by evaluating the intersection point between the mid-plane and the coincidence line detected by the scanner (Fig. 1). The time $t$ was stored along with the spatial coordinates. The advantages of this method with respect to classically used methods in PET imaging, like MLEM, are several. First, it allows to monitor and analyse simultaneously $x$, $y$, $z$ and $t$, allowing for decay rate studies in selected regions. Second, no external image reconstruction programs are needed, making the analysis workflow faster and simpler. 

MC simulations of particle interactions and photon propagation in the phantoms were performed using the FLUKA code~\cite{fluka1,fluka2}. We analysed two distributions: 
\begin{itemize}
\item The activity distributions in space. We investigated the 1-D z-profiles, which is a widely applied method to estimate particle range in patients and to validate the MC simulations. 
\item The activity distributions in time, i.e., the decay rates. We performed an exponential fit to estimate the relative contribution of $^{15}$O ($t_{1/2}$=2 min), $^{11}$C ($t_{1/2}$=20 min) and $^{10}$C ($t_{1/2}$=19 s) in the phantoms, focusing on a time interval from 8 seconds to 5 minutes, where t=0 corresponds to the start of irradiation, and t=5 s to the end of irradiation. The amount of neglected isotopes ($^{5}$B, $^{14}$O, $^{13}$N, ...) was checked with FLUKA to be less than 2\% for the selected time intervals.
\end{itemize}

\section{Results}
In Fig. 3 we display the 1-D activity distribution as predicted by FLUKA for 130 MeV protons on a PMMA target. A good agreement is seen. In fig. 3(d) the ZEBRA structure of the phantom is particularly clear: in the PMMA regions, the activity is much larger than that in the PE regions, because of the contribution of $^{15}$O, which is not produced in PE.
\begin{figure}[b!]
\includegraphics[width=0.8\textwidth]{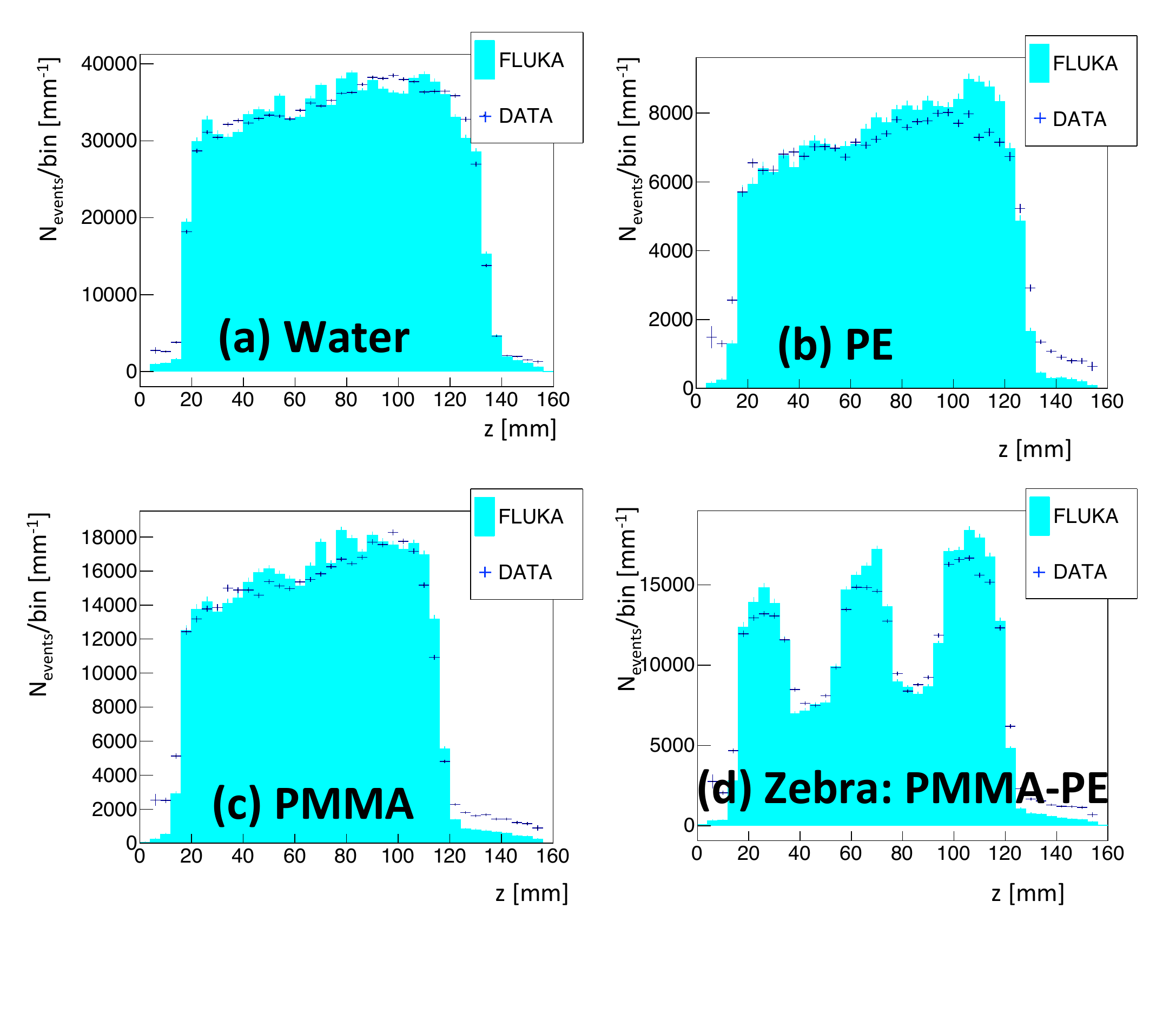}
\caption{\footnotesize{Activity profile along the beam-axis (z). The FLUKA distribution was normalized to the same area as the data. Note that the beam entered the phantom at z=15 mm.}}
\label{fig1}
\end{figure}

In Fig. 4 we display the 1-D decay rate for the four different phantoms fitted with the contributions from $^{15}$O, $^{11}$C, and $^{10}$C in Water (a), PE (b), PMMA (c), and Zebra:PMMA-PE(d). For PE, PMMA and Zebra:PMMA-PE at small times the data are somewhat higher than the FLUKA prediction, however, at large times an excellent agreement is seen. The fitted values were used to calculate the relative fractions of  $^{15}$O, $^{11}$C, and $^{10}$C that decayed in the time interval from 8 to 300 s.
\begin{figure}[t!]
\includegraphics[width=0.8\textwidth]{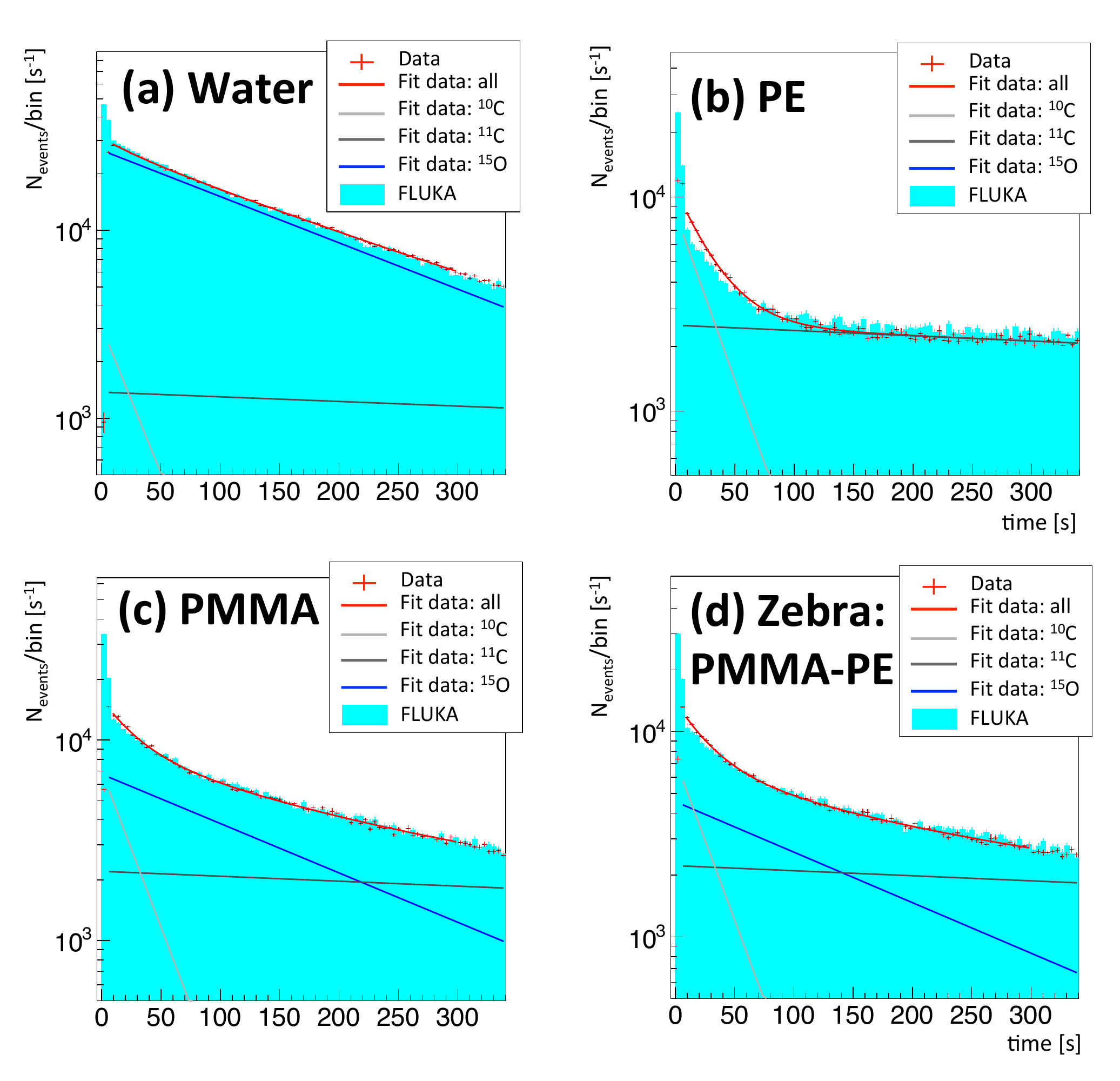}
\caption{\footnotesize{Decay rate as a function of time for the 4 different phantoms, with the contributions from $^{15}O$ (blue), $^{11}$C (dark grey), and $^{10}$C (light grey). The data are displayed in red, and the MC in light blue (filled area).}}
\label{fig4}
\end{figure}

In Tab. 1 we report the MC expected and measured fractions of $^{15}$O, $^{11}$C, and $^{10}$C in Water (a), PE (b), PMMA (c), and Zebra PMMA-PE (d) for a time interval of 8-300 s. The relative fraction of $^{10}$C is in all phantoms somewhat higher in data than in the FLUKA MC simulation. In PE this difference is particularly evident. This is in agreement with earlier reported observations~\cite{cambria}. 
\begin{table}[h!]
\begin{tabular}{|p{2.3cm}|p{1.1cm}|p{1.1cm}|p{1.1cm}|p{1.1cm}|p{1.1cm}|p{1.1cm}|p{1.1cm}|p{1.1cm}|}
\hline
$\beta^+$-emitter & \multicolumn{2}{c|}{Water} &\multicolumn{2}{c|}{PE} & \multicolumn{2}{c|}{PMMA} & \multicolumn{2}{c|}{Zebra: PMMA-PE}\\
\cline{2-9}
& Data& MC & Data & MC & Data & MC & Data & MC\\
\hline
$^{15}$O & 89.4\% & 91.5\%  & 0\%   & 0\%   & 55.5\% & 58.5\% & 45.4\%& 45.7\%\\
\hline
$^{11}$C & 9.1\%  & 6.2\%   & 79.3\%& 85.7\%& 35.7\% & 34.9\% & 43.5\%& 46.2\%\\
\hline
$^{10}$C & 1.6\%  & 2.4\%   & 20.7\%& 14.3\%& 8.8\%  & 6.6\%  & 11.1\% & 8.1\%\\
\hline
\end{tabular}
\caption{\footnotesize{Relative fractions (in \%) of $^{15}$O, $^{11}$C, and $^{10}$C in the time interval from 8 to 300 s. Note that the sum of all columns is 100\%.\label{tab1}}}
\end{table}

By dividing the phantom into different slices (2 mm in $z$, 20 mm in $x$ and $y$) and repeating the fit in each slice, it is possible to approximately map the amount of $^{15}$O, $^{11}$C and $^{10}$C. In Fig. 5 we show the 1-D maps for $^{15}$O, $^{11}$C and $^{10}$C for  the Zebra phantom data in two different time intervals (8-300 s and 8-600 s), for data and MC. We observe the following. First, the different contributions from the various isotopes are different in the PMMA and PE regions. For instance, $^{15}$O is abundantly produced in PMMA, but not at all in the PE regions, as can be seen from the dark blue colour in the $^{15}$O maps. Second, comparing the maps of data acquired in the time interval from 8 to 300 s with that from 8 to 600 s, we see that the maps of $^{15}$O and $^{11}$C are somewhat smoother, due to the increase in statistics. The $^{10}$C map doesn't change, since all $^{10}$C nuclei decay quickly after the end of irradiation ($t_{1/2}$=19 s). Third, by comparing the data and MC maps, we note that the map of $^{10}$C in data (Fig. 5c and 5f) is somewhat different from that in MC (Fig 5i and 5l), especially deeper in the phantom. However, more statistics would be needed to verify this.  

\begin{figure}[t!]
\includegraphics[width=1.1\textwidth]{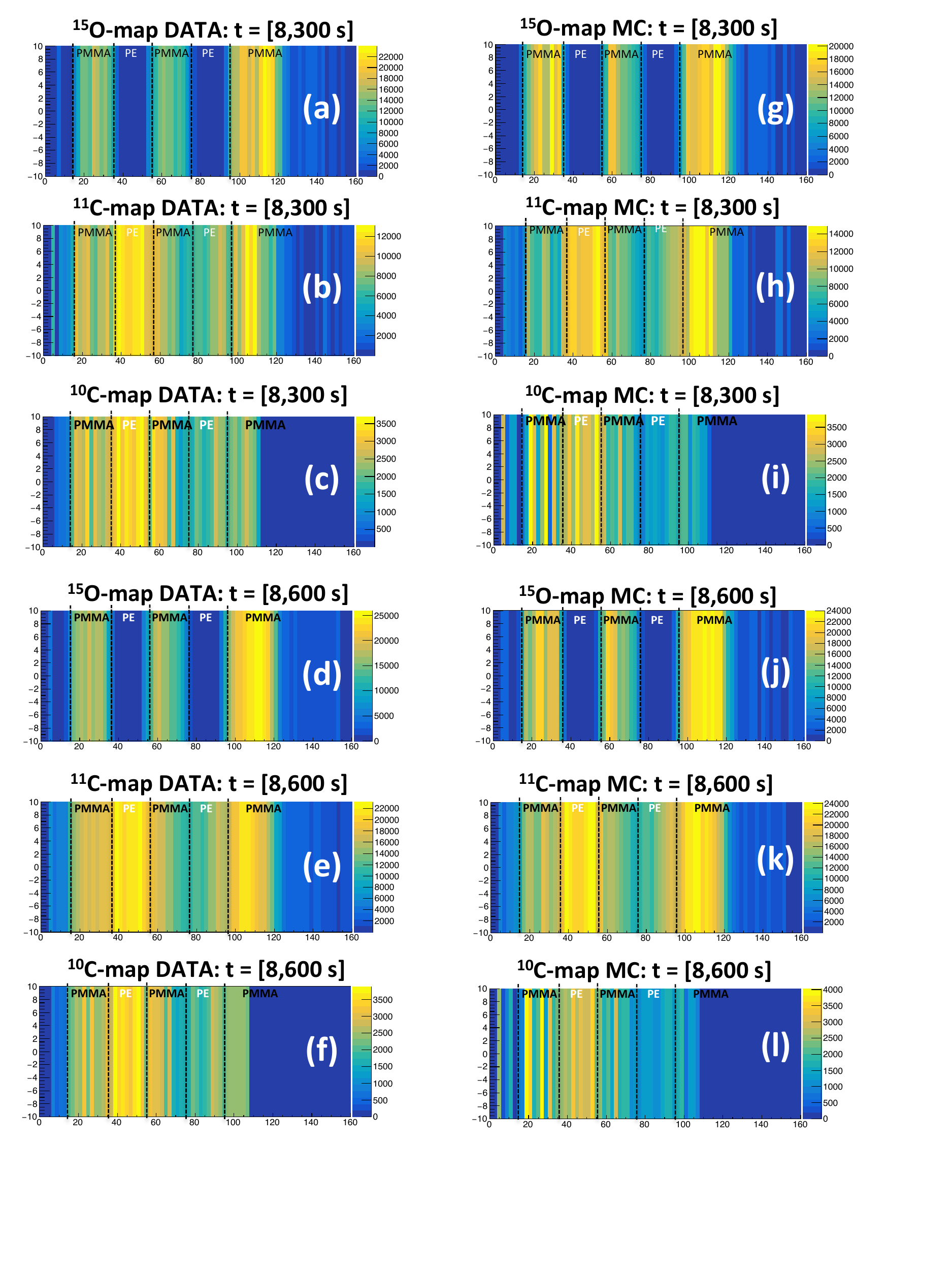}
\vspace*{-3cm}
\caption{\footnotesize{Number of $\beta^+$-decays per slice for $^{15}$O,  $^{11}$C, and $^{10}$C as a function of $z$ for the Zebra phantom, for data in a time-interval from 8 to 300 s (a-c) and from 8 to 600 s (d-f), and MC in a time-interval from 8 to 300 s (g-i) and from 8 to 600 s (j-l).}}

\label{fig1}
\end{figure}

\section{Conclusions:}
We extracted the fractions of $^{15}$O, $^{11}$C and $^{10}$C for various phantoms in time intervals within 5 minutes after irradiation, in the entire phantom and in small slices along the beam. We compared our results with predictions from the FLUKA code and found somewhat more $^{10}$C than what is predicted in FLUKA. This confirms what was  seen in other studies~\cite{cambria}. In the current study we used only 10$^{10}$ protons. With more statistics it would be possible to make more detailed isotope maps, like $^{15}$O, which, if applied in living phantoms, is useful for perfusion studies. The current study is also a starting point for tissue composition studies.



\end{document}